\documentclass[letterpaper,oneside,british]{elsart}
\usepackage{savesym}
\savesymbol{AND}
\usepackage[T1]{fontenc}
\usepackage[latin9]{inputenc}
\usepackage{float}
\usepackage{amsmath}
\usepackage{graphicx}

\makeatletter

\providecommand{\tabularnewline}{\\}
\floatstyle{ruled}
\newfloat{algorithm}{tbp}{loa}
\providecommand{\algorithmname}{Algorithm}
\floatname{algorithm}{\protect\algorithmname}

\usepackage{algorithmic}

\makeatother

\usepackage{babel}
\begin{document}
\begin{frontmatter}

\title{An Iterative Linearised Solution to the Sinusoidal Parameter Estimation
Problem}

\end{frontmatter}
\begin{frontmatter}
\author[ict,xiph]{Jean-Marc Valin\corauthref{cor}},
\ead{jean-marc.valin@csiro.au}
\author[tas]{Daniel V. Smith},
\ead{daniel.v.smith@csiro.au}
\author[rh,xiph]{Christopher Montgomery},
\ead{xiphmont@xiph.org}
\author[xiph]{Timothy B. Terriberry}
\ead{tterribe@xiph.org}
\corauth[cor]{Corresponding author. Address: CSIRO ICT Centre, Cnr. Vimiera \& Pembroke Roads, Marsfield NSW 2122, Australia. Tel: +61 (0)2 9372 4284.}\address[ict]{CSIRO ICT Centre, Australia}
\address[tas]{CSIRO Tasmanian ICT Centre, Australia}
\address[rh]{RedHat Inc., USA}
\address[xiph]{Xiph.Org Foundation}
\begin{abstract}
Signal processing applications use sinusoidal modelling for speech synthesis, speech coding, and audio coding. Estimation of the model parameters involves non-linear optimisation methods, which can be very costly for real-time applications. We propose a low-complexity iterative method that starts from initial frequency estimates and converges rapidly. We show that for $N$ sinusoids in a frame of length $L$, the proposed method has a complexity of $O(LN)$, which is significantly less than the matching pursuits method. Furthermore, the proposed method is shown to be more accurate than the matching pursuits and time-frequency reassignment methods in our experiments. 
\end{abstract}
\end{frontmatter}

\section{Introduction}

Signal processing applications such as speech synthesis~\cite{Stylianou2001},
speech coding~\cite{Hedelin1983}, and audio coding~\cite{LevineThesis}
increasingly use sinusoidal models. Estimating the model parameters
often represents a significant fraction of their overall computational
complexity. Real-time applications require a very low-complexity estimation
algorithm.

This paper proposes a new parameter estimation procedure based on
the linearisation of the model around an initial frequency estimate
and iterative optimisation with fast convergence. For typical configurations,
it is over 20 times less complex than matching pursuits~\cite{mallat93matching}. 

We start by introducing sinusoidal modelling and prior art in Section~\ref{sec:Sinusoidal-Modelling}.
Section~\ref{sec:Frequency-Estimation} discusses frequency estimation
and our proposed linearisation. In Section~\ref{sec:Iterative-Solver},
we present a low-complexity iterative solver for estimating sinusoidal
parameters. Results are discussed in Section~\ref{sec:Results-And-Discussion},
and Section~\ref{sec:Conclusion} concludes this paper. Unless otherwise
noted, a bold uppercase symbol ($\mathbf{A}$) denotes a matrix, a
bold lower case symbol ($\mathbf{a}_{i}$) denotes a column of the
matrix, and an italic symbol ($a_{i,j}$) denotes an element of the
matrix.

\section{Sinusoidal Parameter Estimation\label{sec:Sinusoidal-Modelling}}

A general sinusoidal model that considers both amplitude and frequency
modulation can be used to approximate a signal $\tilde{x}\left(t\right)$
as:
\begin{align}
\tilde{x}\left(t\right) & =\sum_{k=1}^{N}A_{k}\left(t\right)\cos\left(\int_{0}^{t}\omega_{k}\left(u\right)du+\phi_{k}\right)\ ,\label{eq:general-sinusoidal}
\end{align}
where $A_{k}\left(t\right)$ is the time-varying amplitude, $\omega_{k}\left(t\right)$
is the time-varying frequency and $\phi_{k}$ is the initial phase.
The model in~\eqref{eq:general-sinusoidal} has limited practical
use because there are an arbitrary number of ways to approximate $A_{k}\left(t\right)$
and $\omega_{k}\left(t\right)$. Using discrete time $n$ and normalised
frequencies $\theta_{k}$ over a finite window $h\left(n\right)$
yields a simpler model:
\begin{equation}
\tilde{x}\left(n\right)=h\left(n\right)\sum_{k=1}^{N}\left(A_{k}+\dot{A}_{k}n\right)\cos\left(\theta_{k}n+\phi_{k}\right)\ ,\label{eq:4-term-general}
\end{equation}
where $\dot{A}$ is the first time derivative of the amplitude, or
even
\begin{equation}
x\left(n\right)=h\left(n\right)\sum_{k=1}^{N}A_{k}\cos\left(\theta_{k}n+\phi_{k}\right)\ ,\label{eq:3-term-general}
\end{equation}
if we do not want to model amplitude variation within a frame. Although
simpler, the models in~\eqref{eq:4-term-general} and~\eqref{eq:3-term-general}
are still difficult to estimate because they involve a non-linear
optimisation problem.

There are several methods for estimating these sinusoidal model parameters.
The simplest is a standard discrete Fourier transform (DFT) over a
rectangular window. This is limited by frequency leakage caused by
sidelobes from the rectangular window and by its poor frequency resolution\footnote{Throughout this paper, ``resolution'' means the smallest frequency
difference that can be measured for a sinusoid, not the capability
to distinguish between two close sinusoids.}, which is $2\pi/L\ rad/s$ for a frame of length $L$. 

By defining an over-complete dictionary of sinusoidal bases, matching
pursuits methods~\cite{mallat93matching} make it possible to increase
the frequency resolution arbitrarily. Their basis functions also allow
a non-rectangular window to reduce sidelobes. However, as a greedy
algorithm, matching pursuits behaves sub-optimally when the basis
functions are not orthogonal~\cite{vos50hqc}, which is usually the
case for sinusoids of arbitrary frequency over a finite window length.
The orthogonality problem of matching pursuits can mainly be overcome
by further non-linear optimisation as in~\cite{vos50hqc}. However,
this increases complexity significantly, to as high as $O\left(N^{4}\right)$. 

The time-frequency reassignment (TFR) method is another approach that
improves the frequency estimate resolution. When using a spectrogram
representation, phase information from the short-time Fourier transform
(STFT) is exploited to reassign energy from the centre of a spectral
bin $(t,w)$ to its centre of gravity, $(t^{*},w^{*})$~\cite{Auger1995,Plante1998}.
The drawback is that this approach is not well suited to noisy signal
conditions, as energy becomes reassigned to noise dominated regions~\cite{Plante1998}.

Other work, such as~\cite{Stylianou2001,Dhaes2004}, focuses on the
estimation of sinusoidal partials in harmonic signals. While these
are generally low complexity methods, they are not applicable to non-harmonic
signals.

\section{Linearised Model\label{sec:Frequency-Estimation}}

We propose another way to obtain accurate frequency estimates, by
rewriting the sinusoidal model in~\eqref{eq:4-term-general} as
\begin{align}
\tilde{x}\left(n\right) & =h\left(n\right)\sum_{k=1}^{N}\left(A_{k}+n\dot{A}_{k}\right)\cdot\cos\left(\left(\theta_{k}+\Delta\theta_{k}\right)n+\phi_{k}\right)\ ,\label{eq:5-term-general}
\end{align}
where $\theta_{k}$ is an initial estimate of the frequencies and
$\Delta\theta_{k}$ is an unknown correction to the initial estimate.
When both the amplitude modulation parameter $\dot{A}_{k}$ and the
frequency correction $\Delta\theta_{k}$ are small, we show in Appendix
\ref{sec:Linearisation} that~\eqref{eq:5-term-general} can be linearised
as the sum of four basis functions:
\begin{align}
\tilde{x}\left(n\right) & \approx h\left(n\right)\sum_{k=1}^{N}c_{k}\cos\theta_{k}n+s_{k}\sin\theta_{k}n+d_{k}n\cos\theta_{k}n+t_{k}n\sin\theta_{k}n\ ,\label{eq:least-square-prob}
\end{align}
with
\begin{align}
c_{k} & =A_{k}\cos\phi_{k}\ ,\label{eq:lin_ck_param1}\\
s_{k} & =-A_{k}\sin\phi_{k}\ ,\label{eq:lin_sk_param1}\\
d_{k} & =\dot{A}_{k}\cos\phi_{k}-A_{k}\Delta\theta_{k}\sin\phi_{k}\ ,\label{eq:lin_dk_param1}\\
t_{k} & =-\dot{A}_{k}\sin\phi_{k}-A_{k}\Delta\theta_{k}\cos\phi_{k}\ .\label{eq:lin_tk_param1}
\end{align}

\begin{figure}[h]
\begin{center}\includegraphics[width=0.7\columnwidth]{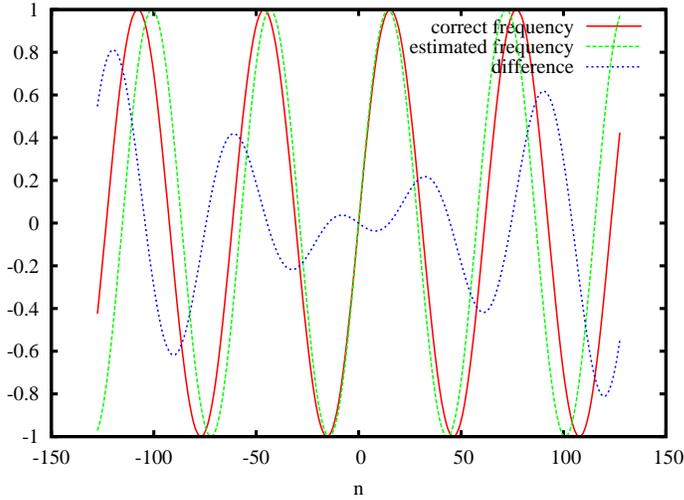}\end{center}

\caption{Difference between two sinusoids of nearly identical frequencies,
resulting in an amplitude-modulated sinusoid.\label{sinusoid-difference}}
\end{figure}

Fig.~\ref{sinusoid-difference} is a visual demonstration of the
linearisation for a small frequency correction. It shows that if the
frequency estimate is very close to the actual frequency of the sinusoid,
the error between the estimated sinusoid and the actual sinusoid can
be approximated as an amplitude modulated sinusoid. Hence, that error
can be modelled using the two basis functions~\eqref{eq:lin_dk_param1}
and~\eqref{eq:lin_tk_param1}.

We can express~\eqref{eq:least-square-prob} in matrix form as
\begin{align}
\tilde{\mathbf{x}} & \approx\mathbf{A}\mathbf{w}\ ,\\
\mathbf{A} & =\left[\mathbf{A}^{c},\mathbf{A}^{s},\mathbf{A}^{d},\mathbf{A}^{t}\right]\ ,\\
\mathbf{w} & =\left[\mathbf{c},\mathbf{s},\mathbf{d},\mathbf{t}\right]^{T}\ ,
\end{align}
where the basis components $\mathbf{A}^{c}$, $\mathbf{A}^{s}$, $\mathbf{A}^{d}$,
and $\mathbf{A}^{t}$ are defined as
\begin{align}
a_{n,k}^{c} & =h\left(n\right)\cos\theta_{k}n\ ,\label{eq:basis_fn1}\\
a_{n,k}^{s} & =h\left(n\right)\sin\theta_{k}n\ ,\\
a_{n,k}^{d} & =h\left(n\right)n\cos\theta_{k}n\ ,\\
a_{n,k}^{t} & =h\left(n\right)n\sin\theta_{k}n\ .\label{eq:basis_fn4}
\end{align}
The best fit is obtained through the least-squares optimisation
\begin{align}
\underset{\mathbf{w}}{\min}\left\Vert \mathbf{A}\mathbf{w}-\mathbf{x}_{h}\right\Vert ^{2} & \ ,
\end{align}
where $\mathbf{x}_{h}$ is the windowed input signal. This leads to
the well known solution
\begin{equation}
\mathbf{w}=\left(\mathbf{A}^{T}\mathbf{A}\right)^{-1}\mathbf{A}^{T}\mathbf{x}_{h}\ .\label{eq:explicit-least-square}
\end{equation}

Once the linear parameters in~\eqref{eq:least-square-prob} are found,
the original sinusoidal parameters can be retrieved by solving the
system~\eqref{eq:lin_ck_param1}-\eqref{eq:lin_tk_param1}:
\begin{align}
A_{k} & =\sqrt{c_{k}^{2}+s_{k}^{2}}\ ,\\
\phi_{k} & =\arg\left(c_{k}-\jmath s_{k}\right)\ ,\\
\dot{A}_{k} & =\frac{d_{k}c_{k}+s_{k}t_{k}}{A_{k}}\ ,\\
\Delta\theta_{k} & =\frac{d_{k}s_{k}-t_{k}c_{k}}{A_{k}^{2}}\ .
\end{align}

\subsection{Frequency Modulation and Higher Order Terms}

Generalising the approach to include second order basis functions
yields
\begin{align}
a_{n,k}^{f} & =h\left(n\right)n^{2}\cos\theta_{k}n\ ,\\
a_{n,k}^{u} & =h\left(n\right)n^{2}\sin\theta_{k}n\ .
\end{align}
This allows the estimation of both the second derivative of the amplitude,
$\ddot{A}$, and the derivative of the frequency, $\dot{\theta}$,
resulting in the following model:
\begin{align}
\tilde{x}\left(n\right) & =h\left(n\right)\sum_{k=1}^{N}\left(A_{k}+n\dot{A}_{k}+n^{2}\ddot{A}_{k}\right)\cdot\cos\left(\left(\theta_{k}+\Delta\theta_{k}+\dot{\theta}_{k}n\right)n+\phi_{k}\right)\ ,
\end{align}
Appendix~\ref{sec:Derivation-2nd-order} derives the second order
linearised model: 
\begin{multline}
\tilde{x}\left(n\right)\approx h\left(n\right)\sum_{k=1}^{N}c_{k}\cos\theta_{k}n+s_{k}\sin\theta_{k}n\\
+d_{k}n\cos\theta_{k}n+t_{k}n\sin\theta_{k}n\ \\
+f_{k}n^{2}\cos\theta_{k}n+u_{k}n^{2}\sin\theta_{k}n\ ,\label{eq:2ndleast-square-prob}
\end{multline}
with
\begin{align}
c_{k} & =A_{k}\cos\phi_{k}\ ,\label{eq:lin_ck_param2}\\
s_{k} & =-A_{k}\sin\phi_{k}\ ,\\
d_{k} & =\dot{A}_{k}\cos\phi_{k}-A_{k}\Delta\theta_{k}\sin\phi_{k}\ ,\\
t_{k} & =-\dot{A}_{k}\sin\phi_{k}-A_{k}\Delta\theta_{k}\cos\phi_{k}\ ,\ \label{eq:lin_tk_param2}\\
f_{k} & =\ddot{A}_{k}\cos\phi_{k}-A_{k}\dot{\theta}_{k}\sin\phi_{k}\ ,\\
u_{k} & =-\ddot{A}_{k}\sin\phi_{k}-A_{k}\dot{\theta}_{k}\cos\phi_{k}\ .\label{eq:lin_uk_param2}
\end{align}
The second order model~\eqref{eq:2ndleast-square-prob} can be formulated
in matrix form: 
\begin{align}
\tilde{\mathbf{x}} & \approx\mathbf{A}\mathbf{w}\ ,\\
\mathbf{A} & =\left[\mathbf{A}^{c},\mathbf{A}^{s},\mathbf{A}^{d},\mathbf{A}^{t},\mathbf{A}^{f},\mathbf{A}^{u}\right]\ ,\\
\mathbf{w} & =\left[\mathbf{c},\mathbf{s},\mathbf{d},\mathbf{t},\mathbf{f},\mathbf{u}\right]^{T}\ ,
\end{align}
where the basis components $\mathbf{A}^{c}$, $\mathbf{A}^{s}$, $\mathbf{A}^{d}$,
$\mathbf{A}^{t}$, $\mathbf{A}^{f}$, and $\mathbf{A}^{u}$ are defined
as
\begin{align}
a_{n,k}^{c}= & h\left(n\right)\cos\theta_{k}n\ ,\label{eq:2ndbasis_fn1}\\
a_{n,k}^{s}= & h\left(n\right)\sin\theta_{k}n\ ,\\
a_{n,k}^{d}= & h\left(n\right)n\cos\theta_{k}n\ ,\\
a_{n,k}^{t}= & h\left(n\right)n\sin\theta_{k}n\ ,\\
a_{n,k}^{f}= & h\left(n\right)n^{2}\cos\theta_{k}n\ ,\\
a_{n,k}^{u}= & h\left(n\right)n^{2}\sin\theta_{k}n\ .
\end{align}

As with the first order model, a least-squares optimisation can be
used to obtain the linear terms \eqref{eq:lin_ck_param2}-\eqref{eq:lin_uk_param2}.
The explicit sinusoidal parameters can then be computed with

\begin{align}
A_{k} & =\sqrt{c_{k}^{2}+s_{k}^{2}}\ ,\\
\phi_{k} & =\arg\left(c_{k}-\jmath s_{k}\right)\ ,\\
\dot{A}_{k} & =\frac{d_{k}c_{k}+s_{k}t_{k}}{A_{k}}\ ,\\
\Delta\theta_{k} & =\frac{d_{k}s_{k}-t_{k}c_{k}}{A_{k}^{2}}\ ,\\
\ddot{A}_{k} & =\frac{f_{k}c_{k}+s_{k}u_{k}}{A_{k}},\ \\
\dot{\theta}_{k} & =\frac{f_{k}s_{k}-u_{k}c_{k}}{A_{k}^{2}},
\end{align}

The first and second order models are identical, apart from the addition
of the $\ddot{A}$ and $\dot{\theta}$ terms, which model quadratic
amplitude modulation and linear frequency modulation, respectively.
The analysis in Appendix \ref{sec:Derivation-2nd-order} makes clear
what the third order model and above would look like. However, the
accuracy of each additional set of terms decreases with the order,
limiting the usefulness of higher order models.

\section{Iterative Solver\label{sec:Iterative-Solver}}

Though solving the linear system~\eqref{eq:explicit-least-square}
demands far less computation than a classic non-linear solver, it
still requires a great amount. D'haes proposed a method that reduces
that complexity from $O\left(LN^{2}\right)$ to $O\left(N\log N\right)$,
but only for harmonic signals~\cite{Dhaes2004}. In this paper, we
propose an $O\left(LN\right)$ solution without the restriction to
harmonic signals.

Our method uses an iterative solution based on the assumption that
matrix $A$ is close to orthogonal, so that
\begin{equation}
\left(\mathbf{A}^{T}\mathbf{A}\right)^{-1}\approx\mathrm{diag}\left\{ \frac{1}{\mathbf{a}_{1}^{T}\mathbf{a}_{1}},\ldots,\frac{1}{\mathbf{a}_{N}^{T}\mathbf{a}_{N}}\right\} =\boldsymbol{\Phi}\ .
\end{equation}
This way, an initial estimate can be computed as
\begin{equation}
\mathbf{w}^{(0)}=\boldsymbol{\Phi}^{-1}\mathbf{A}^{T}\mathbf{x}_{h}\label{eq:Jacobi-initial}
\end{equation}
and then refined as
\begin{align}
\mathbf{w}^{(i+1)} & =\mathbf{w}^{(i)}+\boldsymbol{\Phi}^{-1}\mathbf{A}^{T}\left(\mathbf{x}_{h}-\tilde{\mathbf{x}}^{(i)}\right)\nonumber \\
 & =\mathbf{w}^{(i)}+\boldsymbol{\Phi}^{-1}\mathbf{A}^{T}\left(\mathbf{x}_{h}-\mathbf{A}\mathbf{w}^{(i)}\right)\ .\label{eq:Jacobi-update}
\end{align}

The iterative method described in~\eqref{eq:Jacobi-initial}-\eqref{eq:Jacobi-update}
is strictly equivalent to the Jacobi iterative method. The complexity
of the algorithm is reduced to $O(LMN)$, where $M$ is the number
of iterations required for acceptable convergence. Unfortunately,
while in practise the Jacobi method is stable for most matrices $\mathbf{A}$,
convergence is not guaranteed and depends on the actual frequencies
$\theta_{k}$.

\subsection{Gauss-Seidel Method}

An alternative to the Jacobi method is the Gauss-Seidel method. Its
main advantage is that convergence is guaranteed, since the matrix
$\mathbf{A}^{T}\mathbf{A}$ is symmetric and positive definite~\cite{saad2003ims}.
Since the columns of $\mathbf{A}$ are usually nearly orthogonal,
$\mathbf{A}^{T}\mathbf{A}$ is strongly diagonally dominant, and the
Gauss-Seidel method converges quickly. The linear system can be expressed
as 
\begin{align}
\mathbf{R}\mathbf{w} & =\mathbf{b}\ ,
\end{align}
where
\begin{align}
\mathbf{R} & =\mathbf{A}^{T}\mathbf{A}\ ,\\
\mathbf{b} & =\mathbf{A}^{T}\mathbf{x}_{h}\ .
\end{align}
Assuming $\mathbf{A}$ has been pre-normalised ($\mathbf{a}_{k}^{T}\mathbf{a}_{k}=1,\forall k$),
the Gauss-Seidel algorithm becomes
\begin{align}
w_{k}^{\left(i+1\right)}= & b_{k}-\sum_{j<k}r_{k,j}w_{j}^{\left(i+1\right)}-\sum_{j>k}r_{k,j}w_{j}^{\left(i\right)}\nonumber \\
= & \mathbf{a}_{k}^{T}\mathbf{x}_{h}-\sum_{j<k}\mathbf{a}_{k}^{T}\mathbf{a}_{j}w_{j}^{\left(i+1\right)}-\sum_{j>k}\mathbf{a}_{k}^{T}\mathbf{a}_{j}w_{j}^{\left(i\right)}\nonumber \\
= & w_{k}^{\left(i\right)}+\mathbf{a}_{k}^{T}\mathbf{x}_{h}-\sum_{j<k}\mathbf{a}_{k}^{T}\mathbf{a}_{j}w_{j}^{\left(i+1\right)}-\sum_{j\geq k}\mathbf{a}_{k}^{T}\mathbf{a}_{j}w_{j}^{\left(i\right)}\nonumber \\
= & w_{k}^{\left(i\right)}+\mathbf{a}_{k}^{T}\mathbf{x}_{h}-\mathbf{a}_{k}^{T}\left(\mathbf{A}\tilde{\mathbf{w}_{k}}^{\left(i+1\right)}\right)\nonumber \\
= & w_{k}^{\left(i\right)}+\mathbf{a}_{k}^{T}\left(\mathbf{x}_{h}-\mathbf{A}\tilde{\mathbf{w}_{k}}^{\left(i+1\right)}\right)\ ,\label{eq:GS1-final}
\end{align}
where
\begin{multline}
\tilde{\mathbf{w}_{k}}^{\left(i+1\right)}=\left[w_{0}^{\left(i+1\right)},\ldots,w_{k-1}^{\left(i+1\right)},\right.\\
\left.w_{k}^{\left(i\right)},\ldots,w_{N-1}^{\left(i\right)}\right]^{T}\ .
\end{multline}
We can further simplify the computation of~\eqref{eq:GS1-final}
by noting that only one element of $\tilde{\mathbf{w}_{k}}^{\left(i+1\right)}$
changes for each step. Thus we have
\begin{equation}
w_{k}^{\left(i+1\right)}=w_{k}^{\left(i\right)}+\mathbf{a}_{k}^{T}\mathbf{e}_{k}^{\left(i+1\right)}\ ,
\end{equation}
where $\mathbf{e}_{k}^{\left(i+1\right)}$ is the current error in
the approximation, computed recursively as
\begin{equation}
\mathbf{e}_{k}^{\left(i+1\right)}=\left\{ \begin{array}{ll}
\begin{aligned} & \mathbf{e}_{k-1}^{\left(i+1\right)}-\\
 & \quad\left(w_{k-1}^{\left(i+1\right)}-w_{k-1}^{\left(i\right)}\right)\mathbf{a}_{k-1}
\end{aligned}
 & ,\ k\neq0\\
\mathbf{e}_{N}^{\left(i\right)} & ,\ k=0
\end{array}\right.\ .
\end{equation}
The resulting computation is summarised in Algorithm~\ref{alg:Improved-iterative-algorithm}.
If there is only one iteration, then algorithm~\ref{alg:Improved-iterative-algorithm}
is equivalent to a simplified version of the matching pursuits algorithm,
where the atoms (frequency of the sinusoids) have been pre-selected
before the search. From this point of view, the proposed method relaxes
the orthogonality assumption made by the matching pursuits method.

The main difference from the Jacobi method is that the Gauss-Seidel
method includes partial updates of the error term after each extracted
sinusoid. Convergence follows intuitively from the fact that each
individual step is an exact projection that is guaranteed to decrease
the current error $\mathbf{e}$ --- or at worst leave it constant
if the solution is optimal. Since the error term is updated after
each component $k$, placing the highest-energy terms first speeds
up the optimisation. For this reason, we first update the $\cos\theta_{k}n$
and the $\sin\theta_{k}n$ terms, followed by the $n\cos\theta_{k}n$
and the $n\sin\theta_{k}n$ terms. This usually reduces the number
of iterations required, converging in half as many iterations as sparse
conjugate gradient techniques, such as LSQR~\cite{PaigeSaunders1982},
which cannot take advantage of the diagonal dominance of the system.

\begin{algorithm}
\begin{algorithmic}

\STATE Compute basis functions~\eqref{eq:basis_fn1}-\eqref{eq:basis_fn4}.

\STATE $\mathbf{w}^{(0)}\leftarrow\mathbf{0}$

\STATE $\mathbf{e}\leftarrow\mathbf{x}_{h}$

\FORALL{iteration $i$=1\ldots M}

\FORALL{sinusoid component $k=1\ldots 4N$}

\STATE $\Delta w_{k}^{(i)}\leftarrow\mathbf{a}_{k}^{T}\mathbf{e}$

\STATE $\mathbf{e}\leftarrow\mathbf{e}-\mathbf{a}_{k}\Delta w_{k}^{(i)}$

\STATE $w_{k}^{(i)}\leftarrow w_{k}^{(i-1)}+\Delta w_{k}^{(i)}$

\ENDFOR

\ENDFOR

\FORALL{sinusoid $k=1\ldots N$}

\STATE$A_{k}\leftarrow\sqrt{c_{k}^{2}+s_{k}^{2}}$

\STATE$\phi_{k}\leftarrow\arg\left(c_{k}-\jmath s_{k}\right)$

\STATE$\dot{A}_{k}\leftarrow\frac{d_{k}c_{k}+s_{k}t_{k}}{A_{k}}$

\STATE$\Delta\theta_{k}\leftarrow\frac{d_{k}s_{k}-t_{k}c_{k}}{A_{k}^{2}}$

\ENDFOR

\end{algorithmic}

\caption{Iterative linear optimisation\label{alg:Improved-iterative-algorithm}}
\end{algorithm}

We choose $n=0$ to lie in the centre of the frame in~\eqref{eq:basis_fn1}-\eqref{eq:basis_fn4},
between sample $L/2$ and sample $L/2+1$ if $L$ is even, giving
all the $\mathbf{a}_{k}^{c}$ and $\mathbf{a}_{k}^{t}$ vectors even
symmetry and all the $\mathbf{a}_{k}^{s}$ and $\mathbf{a}_{k}^{d}$
vectors odd symmetry. This leads to the following orthogonality properties:
\begin{align}
\left\langle \mathbf{a}_{k}^{c},\mathbf{a}_{k}^{s}\right\rangle  & =0\ ,\label{eq:ortho1}\\
\left\langle \mathbf{a}_{k}^{c},\mathbf{a}_{k}^{d}\right\rangle  & =0\ ,\\
\left\langle \mathbf{a}_{k}^{t},\mathbf{a}_{k}^{s}\right\rangle  & =0\ ,\\
\left\langle \mathbf{a}_{k}^{t},\mathbf{a}_{k}^{d}\right\rangle  & =0\ .\label{eq:ortho4}
\end{align}
Similar properties hold for the second order basis vectors. Because
the even and odd bases are orthogonal to each other, we optimise them
separately as
\begin{align}
\left[\mathbf{c},\mathbf{t},\mathbf{f}\right]^{T}= & \left(\mathbf{A}^{even}{}^{T}\mathbf{A}^{even}\right)^{-1}\mathbf{A}^{even}{}^{T}\mathbf{x}\ ,\\
\left[\mathbf{s},\mathbf{d},\mathbf{u}\right]^{T}= & \left(\mathbf{A}^{odd}{}^{T}\mathbf{A}^{odd}\right)^{-1}\mathbf{A}^{odd}{}^{T}\mathbf{x}\ ,\\
\mathbf{A}^{even}= & \left[\mathbf{A}^{c},\mathbf{A}^{t},\mathbf{A}^{f}\right]\ ,\\
\mathbf{A}^{odd}= & \left[\mathbf{A}^{s},\mathbf{A}^{d},\mathbf{A}^{u}\right]\ .
\end{align}
Not only does the orthogonality accelerate convergence, but it allows
us to split the error $\mathbf{e}$ into half-length even and odd
components, reducing the complexity of each iteration by half.

\subsection{Non-Linear Optimisation}

If the initial frequency estimates $\theta_{k}^{0}$ are close to
the real frequencies of the sinusoids $\theta_{k}$, then the error
caused by the linearisation~\eqref{eq:least-square-prob} is very
small. In this case, Algorithm~\ref{alg:Improved-iterative-algorithm}
should result in values of $\theta_{k}^{0}+\Delta\theta_{k}$ that
are very close to the real frequencies. However, if the initial estimates
deviate significantly from the real values, then it may be useful
to restart the optimisation with
\[
\theta_{k}\leftarrow\theta_{k}+\alpha\Delta\theta_{k}\ .
\]
where $\alpha$ is the update rate. Typically $\alpha=1$. Repeating
the operation several times, we obtain a non-linear iterative solver
for $A_{k}$, $\theta_{k}$, $\dot{A}_{k}$, and $\phi_{k}$, and
optionally for $\dot{\theta}_{k}$ and $\ddot{A}_{k}$. 

It is not necessary to wait for Algorithm~\ref{alg:Improved-iterative-algorithm}
to converge before updating the frequencies $\theta_{k}$. We can
let both the linear part and the non-linear part of the solution run
simultaneously. To do that, we must first subtract the solution of
the previous iteration before restarting the linear optimisation. 

\begin{algorithm}
\begin{algorithmic}

\STATE$\forall k,\ \theta_{k}=\theta_{k}^{0}$

\STATE$\forall k,\ [A_{k},\phi_{k},\dot{A}_{k},\ddot{A}_{k},\dot{\theta}_{k}]\leftarrow0$

\STATE $\mathbf{w}^{(0)}\leftarrow\mathbf{0}$

\STATE $\mathbf{e}\leftarrow\mathbf{x}_{h}$

\FORALL{non-linear iteration $i$=1\ldots M}

\FORALL{sinusoid $k$}

\STATE$c_{k}\leftarrow A_{k}\cos\phi_{k}$

\STATE$s_{k}\leftarrow-A_{k}\sin\phi_{k}$

\STATE$d_{k}\leftarrow\dot{A}_{k}\cos\phi_{k}$

\STATE$t_{k}\leftarrow-\dot{A}_{k}\sin\phi_{k}$

\STATE$^{\dagger}f_{k}\leftarrow\ddot{A}_{k}\cos\phi_{k}-A_{k}\dot{\theta}_{k}\sin\phi_{k}$

\STATE$^{\dagger}u_{k}\leftarrow-\ddot{A}_{k}\sin\phi_{k}+A_{k}\dot{\theta}_{k}\cos\phi_{k}$

\ENDFOR

\STATE$\mathbf{e}\leftarrow\mathbf{x}-\mathbf{A}\mathbf{w}^{(i-1)}$
(result of the last iteration with updated frequency)

\FORALL{sinusoid component $k=1\ldots 4N$}

\STATE $\Delta w_{k}^{(i)}\leftarrow\mathbf{a}_{k}^{T}\mathbf{e}$

\STATE $\mathbf{e}\leftarrow\mathbf{e}-\mathbf{a}_{k}\Delta w_{k}^{(i)}$

\STATE $w_{k}^{(i)}\leftarrow w_{k}^{(i-1)}+\Delta w_{k}^{(i)}$

\ENDFOR

\FORALL{sinusoid $k=1\ldots N$}

\STATE$A_{k}\leftarrow\sqrt{c_{k}^{2}+s_{k}^{2}}$

\STATE$\phi_{k}\leftarrow\arg\left(c_{k}-\jmath s_{k}\right)$

\STATE$\dot{A}_{k}\leftarrow\frac{d_{k}c_{k}+s_{k}t_{k}}{A_{k}}$

\STATE$\Delta\theta_{k}\leftarrow\frac{d_{k}s_{k}-t_{k}c_{k}}{A_{k}^{2}}$

\STATE$\theta_{k}\leftarrow\theta_{k}+\alpha\Delta\theta_{k}$

\STATE$^{\dagger}\ddot{A}_{k}\leftarrow\frac{fkc_{k}+s_{k}u_{k}}{A_{k}}$

\STATE$^{\dagger}\dot{\theta}_{k}\leftarrow\frac{f_{k}s_{k}-u_{k}c_{k}}{A_{k}^{2}}$

\ENDFOR

\ENDFOR

\end{algorithmic}

\caption{Non-linear iterative optimisation, including the second order terms.
Steps marked with $^{\dagger}$ are only applied for the second order
model.\label{alg:Non-linear-iterative-optimisation}}
\end{algorithm}

The non-linear method we propose is detailed in Algorithm~\ref{alg:Non-linear-iterative-optimisation}
and shares some similarities with the Gauss-Newton method~\cite{GaussNewton}.
However, the reparametrisation in~\eqref{eq:lin_ck_param1}-\eqref{eq:lin_tk_param1}
allows updates to $A_{k}$, $\dot{A}_{k}$, and $\phi_{k}$ to be
incorporated into the linear model immediately when solving the normal
equations. This greatly improves convergence compared to a standard
Gauss-Newton iteration in the original parameters. Just like Algorithm~\ref{alg:Improved-iterative-algorithm},
it is possible to reduce the complexity of Algorithm~\ref{alg:Non-linear-iterative-optimisation}
in half by taking advantage of the even-odd symmetry of the basis
functions.

\section{Results And Discussion\label{sec:Results-And-Discussion}}

In this section, we characterise the proposed algorithm and compare
it to other sinusoidal parameter estimation algorithms. We attempt
to make the comparison as fair as possible despite the fact that the
methods we are comparing do not have exactly the same assumptions
or output. Both the linear and the non-linear versions of the proposed
algorithm are evaluated. For all algorithms, we use a \emph{sine window}:
\begin{equation}
h(n)=\cos\pi\frac{n-\left(L+1\right)/2}{L}\ ,
\end{equation}
so that the result of applying the window to both the input signal
$\mathbf{x}$ and the basis functions $\mathbf{a}_{k}$ is equivalent
to a Hanning analysis window. Unless otherwise noted, we use a frame
length $L=256$.

\subsection{Convergence}

\begin{figure}
\begin{center}\includegraphics[width=0.8\columnwidth]{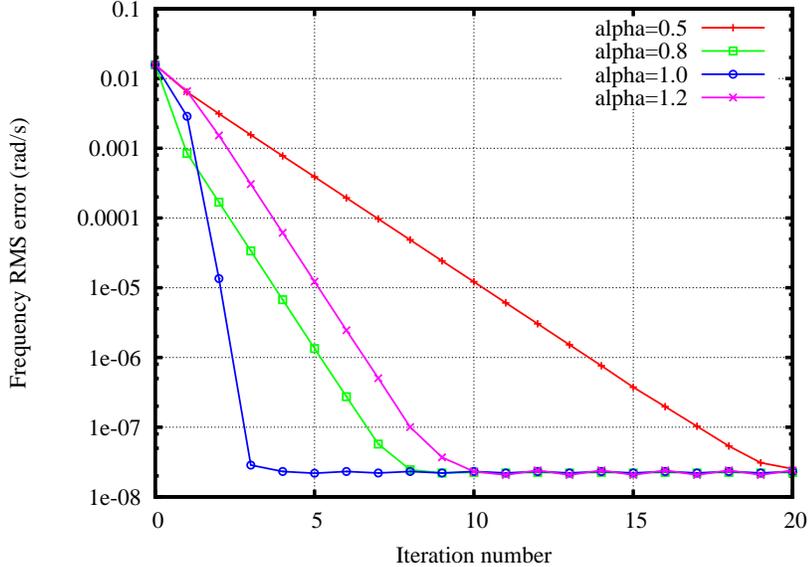}\end{center}

\caption{Convergence of the non-linear optimisation procedure for various values
of $\alpha$. For $\alpha=1$, convergence is achieved in only 3 iterations.
The floor at $2\times10^{-8}\:rad/s$ is due to the finite machine
precision.\label{fig:Convergence-of-nonlinear-optimisation}}
\end{figure}

We first consider the case of a single amplitude-modulated sinusoid
of normalised angular frequency $\theta=0.1\pi$. We start with an
initial frequency estimate of $\theta=0.095\pi$, which corresponds
to an error of slightly more than one period over the 256-sample frames
we use. The non-linear optimisation Algorithm~\ref{alg:Non-linear-iterative-optimisation}
is applied with different values of $\alpha$, using only the first-order
terms. The convergence speed in Figure~\ref{fig:Convergence-of-nonlinear-optimisation}
shows that for $\alpha=1$, convergence becomes much faster than for
other values of $\alpha$, indicating that convergence is super-linear.

If we let the linear part of the algorithm converge at each iteration,
the result is equivalent to the second order Newton's method, since
as shown in Appendix~\ref{sec:Linearisation}, the terms in our linearisation
are equal to a first-order Taylor expansion in the original variables.
Using the chain rule, one can show that Algorithm~\ref{alg:Non-linear-iterative-optimisation}
is only super-linear if the Gauss-Seidel iteration is super-linear.
Since Gauss-Seidel is an iterative linear method, this can only happen
if the basis vectors in $\mathbf{A}$ are orthogonal. In practise,
so long as the separation between frequencies is larger than the sidelobe
of the windowing function, these basis vectors are approximately orthogonal,
although in practise they are never truly orthogonal. However, with
a good choice of windowing function and well-separated frequencies,
convergence is quasi-second order.

If we include second-order terms, then convergence becomes linear,
since the frequency modulation term is not ``recentered'' like the
frequency is. While such recentring is possible, it unnecessarily
increases the complexity of the algorithm while making it more susceptible
to numerical errors.

\begin{figure}
\begin{center}\includegraphics[width=0.8\columnwidth]{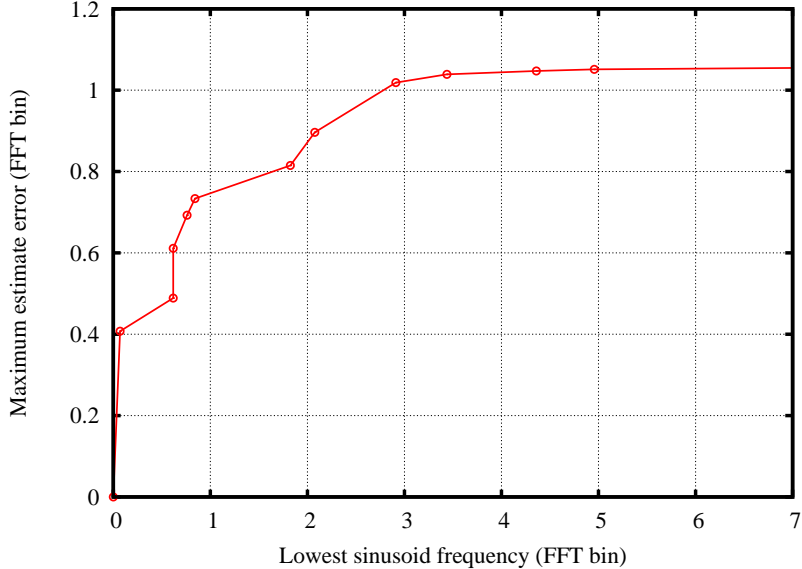}\end{center}

\caption{Region of convergence as a function of the sinusoid frequency. The
algorithm never converges when the initial estimate is off by more
than 1.05 DFT bins ($2\pi/L\,rad/s$ ).\label{fig:Region-of-convergence}}
\end{figure}

As stated in Section~\ref{sec:Frequency-Estimation}, the proposed
algorithm depends on an initial approximation sufficiently close to
the true frequency of a sinusoid. Fig.~\ref{fig:Region-of-convergence}
shows the maximum error in the initial estimate for which the non-linear
algorithm converges to the true frequency. For most frequencies, that
maximum error is equivalent to 1.05 DFT bins. However, for low frequencies,
the tolerance to error is reduced. This is due to the fact that $a_{n,k}^{c}$
becomes highly correlated with $a_{n,k}^{t}$ and $a_{n,k}^{s}$ becomes
highly correlated with $a_{n,k}^{d}$, making it harder to estimate
the frequency offsets $\Delta\theta_{k}$ accurately.

\subsection{Chirps}

Next, we measure the frequency estimation accuracy and the energy
of the residual signal for known signals. We use a synthetic signal
that is the sum of five chirps with white Gaussian noise. The chirps
have linear frequency variations starting at $0.05$, $0.1$, $0.15$,
$0.2$, and $0.25\:rad/s$ and ending at $2.0$, $2.2$, $2.4$, $2.6$,
and $2.8\:rad/s$, respectively. The relative amplitudes of the chirps
are 0 dB, -3 dB, -6 dB, -9 dB, and -12 dB. We consider the following
algorithms:
\begin{itemize}
\item Time-frequency reassignment (\textbf{TFR}),
\item Matching pursuits (32x over-sampled dictionary) (\textbf{MP}),
\item Proposed algorithm with linear optimisation (\textbf{linear}),
\item Proposed algorithm with non-linear optimisation (\textbf{non-linear}),
and
\item Proposed algorithm with non-linear optimisation and second order model
(\textbf{second order}).
\end{itemize}
The time-frequency reassignment method is implemented as in~\cite{Auger1995}.
The matching pursuits algorithm uses a dictionary of non-modulated
sinusoids with a resolution of $\pi/8192$. We also compare to the
theoretical resolution obtained from the picking the highest peaks
in the DFT. These are used as the initial seeds for our algorithm
and TFR. To make sure that algorithms are compared fairly, all algorithms
are constrained to frequencies within one DFT bin of the initial seed,
i.e. there are no outliers. MP does not consider any dictionary elements
outside this range, and any step by the optimisation algorithms is
clamped to lie within it. This occurs only rarely when the SNR is
low.

\begin{figure}
\begin{center}\includegraphics[width=0.8\columnwidth]{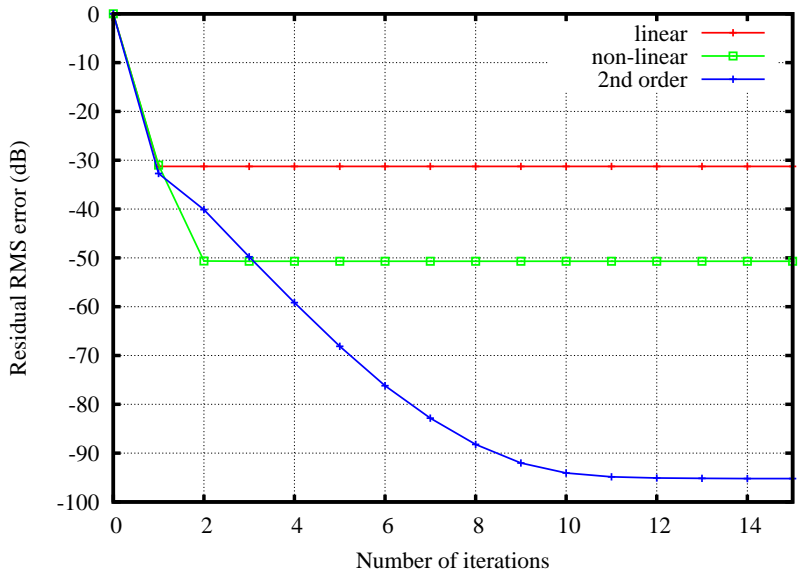}\end{center}

\caption{Reconstruction RMS error as a function of the number of iterations
in clean conditions (linear vs. non-linear)\label{fig:Reconstruction-convergence}}
\end{figure}

Fig.~\ref{fig:Reconstruction-convergence} shows the RMS energy of
the residual ($\tilde{\mathbf{x}}-\mathbf{x}_{h}$) as a function
of the number of iterations for both the linear optimisation and the
non-linear optimisation. The linear version converges after only 2
iterations, while the non-linear version requires 3 iterations. These
are the iteration limits we use for the experiments that follow. In
the case of the second order non-linear version, the convergence continues
until limited by numerical precision, so we limit it to 5 iterations,
which already significantly improves on the first order model.

\begin{figure}
\begin{center}\includegraphics[width=0.8\columnwidth]{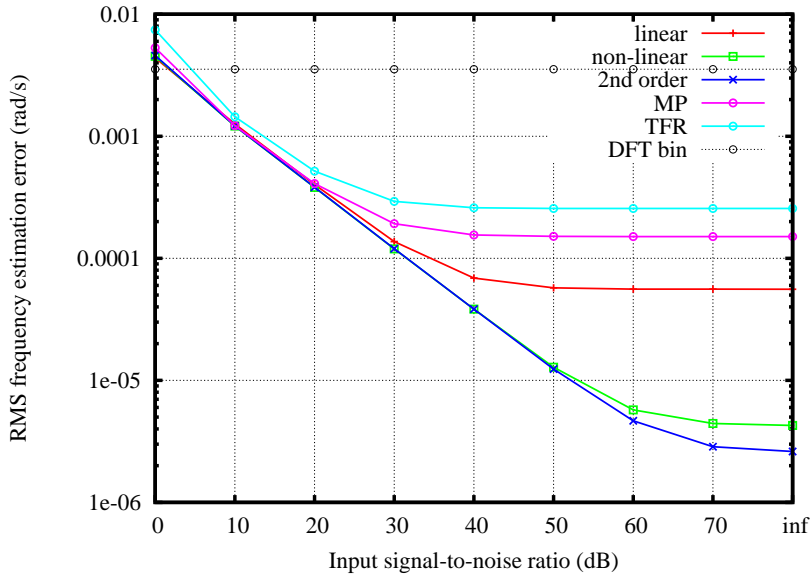}\end{center}

\caption{Frequency RMS estimation error as a function of the SNR.\label{fig:Frequency-RMS-error}}
\end{figure}

Fig.~\ref{fig:Frequency-RMS-error} shows the frequency RMS estimation
error as a function of the SNR for each of the four algorithms. At
very low SNR, all algorithms perform similarly. However, as the SNR
increases above 20 dB, matching pursuits stops improving. This is
likely due to the fact that the frequencies are not orthogonal, which
makes its greedy approach sub-optimal. Both the proposed linear and
non-linear approaches provide roughly the same accuracy up to 30 dB,
after which the non-linear approach provides superior performance.
For this scenario, the only limitation of the non-linear algorithm
at infinite SNR is the fact that it does not account for frequency
modulation within a frame.

\begin{figure}
\begin{center}\includegraphics[width=0.8\columnwidth]{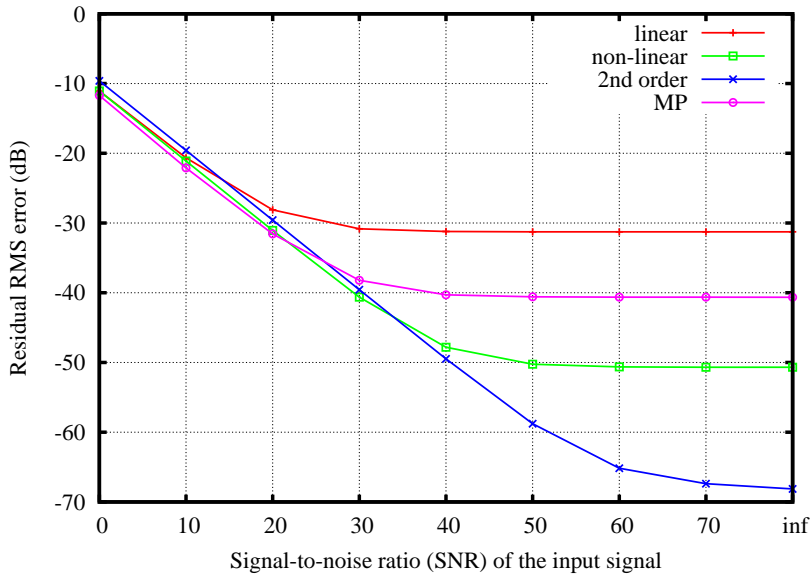}\end{center}

\caption{Reconstruction RMS error as a function of the SNR (the input noise
is not considered in the error).\label{fig:Reconstruction-RMS-error}}
\end{figure}

Fig.~\ref{fig:Reconstruction-RMS-error} shows the reconstruction
error for all algorithms except the time-frequency reassignment method,
which cannot estimate the amplitude and thus cannot provide a reconstructed
signal. The reconstruction error is measured against the noise-free
version of the chirps. The performance mirrors that of Fig.~\ref{fig:Frequency-RMS-error},
with the notable exception that the non-linear optimisation's reconstruction
error plateaus long before the second order method, even though it
is able to accurately estimate the frequency. 

The performance of our algorithm is slightly worse than matching pursuits
at low SNR. This is caused by some slight over-fitting due to the
inclusion of an amplitude modulation term. The difference disappears
if this term is forced to zero.

In the chirp experiments our proposed non-linear algorithms out-perform
both matching pursuits and time-frequency reassignment overall. The
linear version has performance similar to previous methods, but it
does not perform as well as non-linear optimisation. In all cases
(Fig.~\ref{fig:Frequency-RMS-error} to Fig.~\ref{fig:Reconstruction-RMS-error}),
all the algorithms behave similarly. Their error at low SNR is similar,
and the slope of the error curve is the same. The main differentiator
between algorithms is how far they improve with SNR before reaching
a plateau.

\subsection{Audio}

We apply our proposed algorithm to a 90-second collage of diverse
music clips sampled at 48 kHz, including percussive, musical, and
amusical content. In this case, we cannot compare to matching pursuits
because the lack of ground truth prevents us from forcing a common
set of initial sinusoid frequencies. We select the initial frequency
estimates required for the proposed algorithm using peaks in the standard
DFT. The number of sinusoids is variable (depends on the number of
peaks) and a 256-sample window is used.

\begin{figure}
\begin{center}\includegraphics[width=0.8\columnwidth]{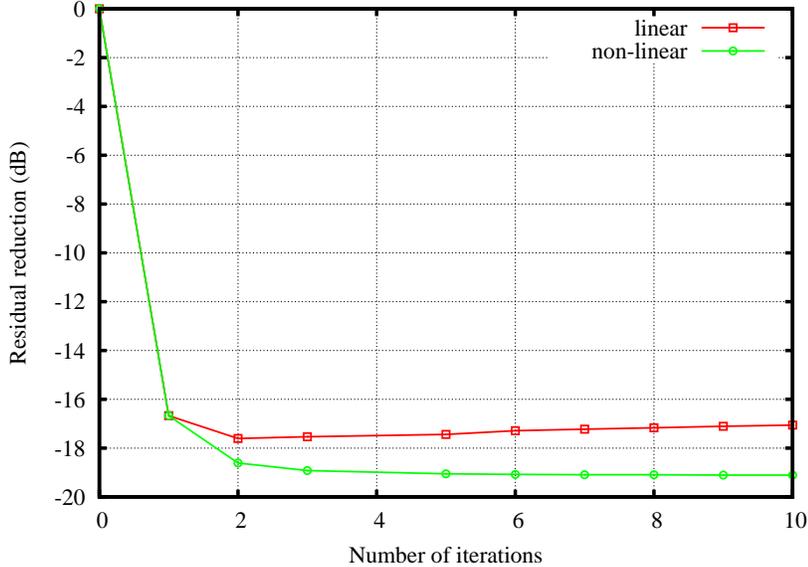}\end{center}

\caption{Reduction in residual energy as a function of the number of iterations.\label{fig:Audio-Reduction-in-residual}}
\end{figure}

The energy of the residual is plotted as a function of the number
of iterations in Fig.~\ref{fig:Audio-Reduction-in-residual}. Both
algorithms converge quickly and we can see that the linear optimisation
only requires 2 iterations, while the non-linear optimisation requires
3 iterations. 

\subsection{Algorithm complexity}

In this section, we compare the complexity of the proposed algorithms
to that of other similar algorithms. For the sake of simplicity, we
discard some terms that are deemed negligible, e.g., we discard $O\left(LN\right)$
terms when $O\left(LN^{2}\right)$ terms are present.

In Algorithm~\ref{alg:Improved-iterative-algorithm}, we can see
that each iteration requires $8LN$ multiplications and $8LN$ additions.
Additionally, computation of the $4N$ basis functions $\mathbf{a}_{k}$
prior to the optimisation requires $LN$ additions and $3LN$ multiplications.
It is possible to further reduce the complexity of each iteration
by taking advantage of the fact that all of our basis functions have
either even or odd symmetry. By decomposing the residual into half-length
even and odd components, only one of these components needs to be
updated for a given basis function. This reduces the complexity of
each iteration in Algorithm~\ref{alg:Improved-iterative-algorithm}
by half without changing the result. The complexity of each iteration
is thus $4LN$ multiplications and $4LN$ additions. For $M$ iterations,
this amounts to a total of $\left(8M+5\right)LN$ operations per frame. 

The complexity of the proposed non-linear optimisation algorithm (Algorithm~\ref{alg:Non-linear-iterative-optimisation})
is similar to that of the linear version, with two exceptions. First,
because the frequencies change every iteration, the basis functions
need to be re-computed each time. Second, when starting a new iteration,
the residual must be updated using the new basis functions. The total
complexity is thus $\left(17M-4\right)LN$ operations per frame. For
a single iteration, the linear and non-linear versions are strictly
equivalent.

By comparison, a simple matching pursuits algorithm that does not
consider modulation requires $2LN^{2}P$ operations per frame, where
$P$ is the oversampling factor, i.e. the increase over the standard
DFT resolution. Using a fast FFT-based implementation~\cite{vos50hqc}
reduces the complexity to $5/2LNP\log_{2}LP$. 

Table~\ref{tab:Complexity-comparison} summarises the complexity
of several algorithms. Because the algorithms have different dependencies
on all the parameters, we also consider the total complexity in Mflops
for real-time estimation of sinusoids in a \emph{typical} scenario,
where we have
\begin{itemize}
\item frame length: $L=256$,
\item number of sinusoids: $N=20$,
\item oversampling: $P=64$ (matching pursuits only),
\item number of iterations: $M=2$ (linear), $M=3$ (non-linear), $M=5$
($2^{nd}$ order)
\item sampling rate: 16 kHz,
\item frame offset: 192 samples (25\% overlap).
\end{itemize}
Table~\ref{tab:Complexity-comparison} shows that the proposed algorithms,
both linear and nonlinear, reduce the complexity by more than an order
of magnitude when compared to matching pursuits algorithms. However,
while matching pursuits can estimate the sinusoidal parameters directly
from the input signal, the proposed method requires initial frequency
estimates. The cost of producing these estimates is not included in
the table but is generally small (e.g., 0.4 Mflops for performing
an FFT).

\begin{table}
\caption{Complexity comparison of various parameter estimation algorithms.
$^{*}$The typical complexity of~\cite{vos50hqc} is not given, but
we estimate it to be at least 500 Mflops, probably much higher. \label{tab:Complexity-comparison} }
\begin{center}%
\begin{tabular}{ccc}
\hline 
Algorithm & Complexity & Typical (Mflops)\tabularnewline
\hline 
Matching pursuits (direct) & $2L^{2}NP$ & 14,000\tabularnewline
Matching pursuits (FFT-based) & $\frac{5}{2}LNP\log_{2}LP$ & 960\tabularnewline
Direct non-linear (\cite{vos50hqc}) & $O\left(N^{4}+LN^{2}\right)$ & >500$^{*}$\tabularnewline
Proposed (linear) & $\left(8M+5\right)LN$ & 9\tabularnewline
Proposed (non-linear) & $\left(17M-4\right)LN$ & 20\tabularnewline
Proposed ($2^{nd}$ order) & $\left(24M-6\right)LN$ & 49\tabularnewline
\hline 
\end{tabular}\end{center}
\end{table}

\section{Conclusion\label{sec:Conclusion}}

We have presented a method for estimating sinusoidal parameters with
very low complexity. It is based on a linearisation of the sinusoidal
model followed by an iterative optimisation of the parameters. The
algorithm converges quickly, requiring only 2 iterations for the linear
optimisation and 3 iterations for the non-linear optimisation. We
showed that the frequency estimation of the non-linear version of
our algorithm is more accurate than the matching pursuits and time-frequency
reassignment methods. In addition, we demonstrated computational complexities
considerably lower than matching pursuits. For applications that require
it, we have also proposed a second order algorithm that estimates
the frequency modulation within a frame. The total complexity of our
approach is more than an order of magnitude less complex than other
proposed methods for estimating sinusoid parameters. Consequently,
our approach could offer significant benefits to areas such as audio
and speech coding, which require sinusoidal modeling to be performed
in real time.

Like other non-linear optimisation methods, ours requires a good initial
estimate of the sinusoids' frequencies. Therefore, low-complexity
sinusoid selection is an important area of future work. 

\appendix

\section{Linearisation of the Sinusoidal Model\label{sec:Linearisation}}

Consider a sinusoidal model with piecewise linear amplitude modulation
and a frequency offset from an initial estimate:
\begin{align}
\tilde{x}\left(n\right) & =\sum_{k=1}^{N}\left(A_{k}+n\dot{A}_{k}\right)\cdot\cos\left(\left(\theta_{k}+\Delta\theta_{k}\right)n+\phi_{k}\right)\ ,
\end{align}
where $\theta_{k}$ is known in advance and $\Delta\theta_{k}$ is
considered small. Using trigonometric identities, we can expand the
sum in the cosine term into
\begin{align}
\tilde{x}\left(n\right) & =\sum_{k=1}^{N}\left(A_{k}+n\dot{A}_{k}\right)\cos\phi_{k}\cos\left(\theta_{k}+\Delta\theta_{k}\right)n\nonumber \\
 & -\sum_{k=1}^{N}\left(A_{k}+n\dot{A}_{k}\right)\sin\phi_{k}\sin\left(\theta_{k}+\Delta\theta_{k}\right)n\\
 & =\sum_{k=1}^{N}\left(A_{k}+n\dot{A}_{k}\right)\cos\phi_{k}\cos\Delta\theta_{k}n\cos\theta_{k}n\nonumber \\
 & -\sum_{k=1}^{N}\left(A_{k}+n\dot{A}_{k}\right)\cos\phi_{k}\sin\Delta\theta_{k}n\sin\theta_{k}n\nonumber \\
 & -\sum_{k=1}^{N}\left(A_{k}+n\dot{A}_{k}\right)\sin\phi_{k}\cos\Delta\theta_{k}n\sin\theta_{k}n\nonumber \\
 & -\sum_{k=1}^{N}\left(A_{k}+n\dot{A}_{k}\right)\sin\phi_{k}\sin\Delta\theta_{k}n\cos\theta_{k}n\ .\label{eq:4term-sinusoidal-full-expand}
\end{align}

In the linearisation process, we further assume that $\Delta\theta_{k}n\ll1$
and $\dot{A}_{k}n\ll A_{k}$, so we can neglect all terms second order
and above. This translates into the following approximations:
\begin{align}
\sin\Delta\theta_{k}n & \approx\Delta\theta_{k}n\ ,\\
\cos\Delta\theta_{k}n & \approx1\ ,\\
n\dot{A}_{k}\sin\Delta\theta_{k}n & \approx0\ .
\end{align}
When substituting the above approximations into~\eqref{eq:4term-sinusoidal-full-expand},
we obtain
\begin{align}
\tilde{x}\left(n\right) & \approx\sum_{k=1}^{N}\left(A_{k}+n\dot{A}_{k}\right)\cos\phi_{k}\cos\theta_{k}n\nonumber \\
 & -\sum_{k=1}^{N}A_{k}\cos\phi_{k}\Delta\theta_{k}n\sin\theta_{k}n\nonumber \\
 & -\sum_{k=1}^{N}\left(A_{k}+n\dot{A}_{k}\right)\sin\phi_{k}\sin\theta_{k}n\nonumber \\
 & -\sum_{k=1}^{N}A_{k}\sin\phi_{k}\Delta\theta_{k}n\cos\theta_{k}n\ .\label{eq:4term-sinusoidal-gather}
\end{align}
Reordering the terms in~\eqref{eq:4term-sinusoidal-gather} leads
to the following formulation:
\begin{align}
\tilde{x}\left(n\right) & \approx\sum_{k=1}^{N}A_{k}\cos\phi_{k}\cos\theta_{k}n\nonumber \\
 & -\sum_{k=1}^{N}A_{k}\sin\phi_{k}\sin\theta_{k}n\nonumber \\
 & +\sum_{k=1}^{N}\left(\dot{A}_{k}\cos\phi_{k}-A_{k}\Delta\theta_{k}\sin\phi_{k}\right)n\cos\theta_{k}n\nonumber \\
 & -\sum_{k=1}^{N}\left(\dot{A}_{k}\sin\phi_{k}+A_{k}\Delta\theta_{k}\cos\phi_{k}\right)n\sin\theta_{k}n\ ,\label{eq:Taylor-equivalent}
\end{align}
which is a linear combination of four functions. The result in~\eqref{eq:Taylor-equivalent}
is in fact equivalent to a first-order Taylor expansion.

\section{Derivation For the Second Order Model\label{sec:Derivation-2nd-order}}

Keeping second order terms allows us to model both the first derivative
of the frequency and the second derivative of the amplitude with respect
to time:

\begin{align}
\tilde{x}\left(n\right) & =\sum_{k=1}^{N}\left(A_{k}+\dot{A}_{k}n+n^{2}\ddot{A}_{k}\right)\cos\phi_{k}\cos\left(\theta_{k}+\Delta\theta_{k}+\dot{\theta}_{k}n\right)n\nonumber \\
 & -\sum_{k=1}^{N}\left(A_{k}+\dot{A}_{k}n+\ddot{A}_{k}n^{2}\right)\sin\phi_{k}\sin\left(\theta_{k}+\Delta\theta_{k}+\dot{\theta}_{k}n\right)n\\
 & =\sum_{k=1}^{N}\left(A_{k}+\dot{A}_{k}n+\ddot{A}_{k}n^{2}\right)\cos\phi_{k}\cos\left(\Delta\theta_{k}n+\dot{\theta}_{k}n^{2}\right)\cos\theta_{k}n\nonumber \\
 & -\sum_{k=1}^{N}\left(A_{k}+\dot{A}_{k}n+\ddot{A}_{k}n^{2}\right)\cos\phi_{k}\sin\left(\Delta\theta_{k}n+\dot{\theta}_{k}n^{2}\right)\sin\theta_{k}n\nonumber \\
 & -\sum_{k=1}^{N}\left(A_{k}+\dot{A}_{k}n+\ddot{A}_{k}n^{2}\right)\sin\phi_{k}\cos\left(\Delta\theta_{k}n+\dot{\theta}_{k}n^{2}\right)\sin\theta_{k}n\nonumber \\
 & -\sum_{k=1}^{N}\left(A_{k}+\dot{A}_{k}n+\ddot{A}_{k}n^{2}\right)\sin\phi_{k}\sin\left(\Delta\theta_{k}n+\dot{\theta}_{k}n^{2}\right)\cos\theta_{k}n\ .\label{eq:2nd4term-sinusoidal-full-expand}
\end{align}

This time, we neglect third order terms in $n$. Non-linear terms
involving the parameters (e.g. $\dot{A}_{k}\Delta\theta_{k}$) are
discarded as well. This leads to
\begin{align}
\sin\left(\Delta\theta_{k}n+\dot{\theta}_{k}n^{2}\right) & \approx\Delta\theta_{k}n\ +\dot{\theta}_{k}n^{2}\ ,\\
\cos\left(\Delta\theta_{k}n+\dot{\theta}_{k}n^{2}\right) & \approx1\ ,\\
\left(\dot{A}_{k}n+\ddot{A}_{k}n^{2}\right)\sin\left(\Delta\theta_{k}n+\dot{\theta}_{k}n^{2}\right) & \approx0\ ,
\end{align}
Substituting into \eqref{eq:2nd4term-sinusoidal-full-expand}, we
obtain
\begin{align}
\tilde{x}\left(n\right) & \approx\sum_{k=1}^{N}A_{k}\cos\phi_{k}\cos\theta_{k}n\nonumber \\
 & -\sum_{k=1}^{N}A_{k}\sin\phi_{k}\sin\theta_{k}n\nonumber \\
 & +\sum_{k=1}^{N}\left(\dot{A}_{k}\cos\phi_{k}-A_{k}\Delta\theta_{k}\sin\phi_{k}\right)n\cos\theta_{k}n\nonumber \\
 & -\sum_{k=1}^{N}\left(\dot{A}_{k}\sin\phi_{k}+A_{k}\Delta\theta_{k}\cos\phi_{k}\right)n\sin\theta_{k}n\ \nonumber \\
 & +\sum_{k=1}^{N}\left(\ddot{A}_{k}\cos\phi_{k}-A_{k}\dot{\theta}_{k}\sin\phi_{k}\right)n^{2}\cos\theta_{k}n\nonumber \\
 & -\sum_{k=1}^{N}\left(\ddot{A}_{k}\sin\phi_{k}+A_{k}\dot{\theta}_{k}\cos\phi_{k}\right)n^{2}\sin\theta_{k}n\ .
\end{align}

\bibliographystyle{unsrt}
\bibliography{sinusoids}

\end{document}